\documentclass[aps,pra,twocolumn,amsmath,amssymb,nofootinbib]{revtex4-1}
\usepackage{graphicx}
\usepackage{dcolumn}
\usepackage{braket}
\usepackage{setspace}
\usepackage{bm}
\usepackage{amsmath}
\usepackage{csquotes}
\usepackage{mathrsfs}
\usepackage{subfigure}
\usepackage[svgnames]{xcolor}
\usepackage[colorlinks,citecolor=red,urlcolor=red,linkcolor=red,bookmarks=false, hypertexnames=true]{hyperref}

\begin{document}
\title{{Noisy Coherent Population Trapping:\\ 
Applications to Noise Estimation and Qubit State Preparation}}
\author{Arshag Danageozian}
\email{Corresponding Author: arshag.danageozian@gmail.com}
\author{Ashe Miller}
\author{Pratik J. Barge}
\author{Narayan Bhusal}
\author{Jonathan P. Dowling}
\thanks{Deceased, June 5 2020.}
\affiliation{Hearne Institute for Theoretical Physics and Department of Physics \& Astronomy, Louisiana State University, Baton Rouge, LA 70803, USA}

\date{\today}
\begin{abstract}
Coherent population trapping is a well-known quantum phenomenon in a driven $\Lambda$ system, with many applications across quantum optics. However, when a stochastic bath is present in addition to vacuum noise, the observed trapping is no longer perfect. Here we derive a time-convolutionless master equation describing the equilibration of the $\Lambda$ system in the presence of additional temporally correlated classical noise, with an unknown decay parameter. Our simulations show a one-to-one correspondence between the decay parameter and the depth of the characteristic dip in the photoluminescence spectrum, thereby enabling the unknown parameter to be estimated from the observed spectra. We apply our analysis to the problem of qubit state initialization in a $\Lambda$ system via dark states and show how the stochastic bath affects the fidelity of such initialization as a function of the desired dark-state amplitudes. We show that an optimum choice of Rabi frequencies is possible.
\end{abstract}

\maketitle

\section{Introduction}

Coherent population trapping (CPT) \cite{alzetta1997induced, scully1997quantum, fleischhauer2005electromagnetically, bergmann1998coherent, arimondo1996v, dalton1982effects, agap1993coherent} is a quantum mechanical phenomenon in driven three-level $\Lambda$ systems used to make a specific material transparent to certain frequencies. Under appropriate driving conditions, the dynamics of the $\Lambda$ system gets ``trapped'' into the Hilbert subspace of the two ground levels, in a coherent superposition which can no longer absorb the light. Such a superposition is known as a ``dark state,'' because it is no longer coupled to the excited state and fluorescent light emission is then suppressed. Under current advances in quantum control, applications of CPT have attracted growing interest outside the field of optics. In the context of dissipative quantum state preparation \cite{hilser2012all,ticozzi2012hamiltonian,yale2013all,pingault2014all,chu2015all,zhou2017dark}, this concept is used to stabilize arbitrary linear superpositions of two ground states by driving the $\Lambda$ system into the (unique) dark state, with the amplitudes of the superposition being determined by the ratio between the two Rabi frequencies and the relative phase between the two laser fields. Notably, CPT plays an important role in protocols for all-optical manipulations in nitrogen-vacancy (NV) centers in diamond \cite{santori2006coherent0,santori2006coherent,golter2013nuclear,jamonneau2016coherent}. More recently, CPT has found application in real-time quantum sensing, by allowing the effective magnetic field in a medium to be estimated via the rate of photon counts under CPT conditions
\cite{WangCPTsensing}. 

Currently, standard theoretical analyses of CPT only account for decoherence due to the quantum vacuum \cite{scully1997quantum,qi2009electromagnetically, whitley1976double}. However, this need not be the only source of noise in many realistic settings of interest. Even assuming that any operational source of noise (e.g., control amplitude or frequency fluctuations) may be experimentally minimized, it is important to expand the treatment to include noise arising directly from the hosting medium in which the $\Lambda$ system is implemented. While we can argue for a noise model that is specific for each medium (environment) on physical grounds, the resulting functional forms will typically still have unknown (e.g., decay) noise parameters that need to be estimated from experimentally accessible quantities. 

In this work, after describing the physical setting in Sec.~\ref{setting}, we theoretically analyze the CPT dynamics of a general $\Lambda$ system under the simultaneous presence of vacuum noise and noise due a classical stochastic environment (Sec.~\ref{NCPT}). Our approach is based on deriving an appropriate time-convolutionless (TCL) master equation (ME) \cite{breuer2002theory}. Based on our analysis, we first show  (Sec.~\ref{applications1}) a correspondence between the height of the dip in the CPT photoluminescence spectrum and the unknown decay parameter of the classical environment, thereby enabling an estimation of this parameter from observed spectra. In Sec.~\ref{applications2}, we further apply this result to quantify the fidelity loss that the noise induces in CPT-based dissipative state initialization, as considered in \cite{yale2013all}. Thus, in our analysis, CPT serves two different but complementary purposes: decay parameter estimation and dissipative state preparation.

While our theoretical approach may be applied to an arbitrary $\Lambda$ system in principle, we use the NV center \cite{tamarat2008spin,chu2015quantum, childress2013diamond} as a realistic illustrative setting for our analysis. 
NV centers are highly studied solid-state systems due to both their long qubit coherence times (ranging from $10^{-6}$s to $10^{-3}$s depending on the isotopic purity of the diamond sample  \cite{maurer2012room,doherty2013nitrogen,kennedy2002single}) and their dynamic accessibility for initialization and read-out using optical pulses. Furthermore, they are scalable solid-state systems \cite{bernien2013heralded}, which makes them a good candidate for various quantum technology applications. 

\section{Physical setting}
\label{setting}

\begin{figure*}[!t] 
    \centering
    \includegraphics[width=1.0\textwidth]{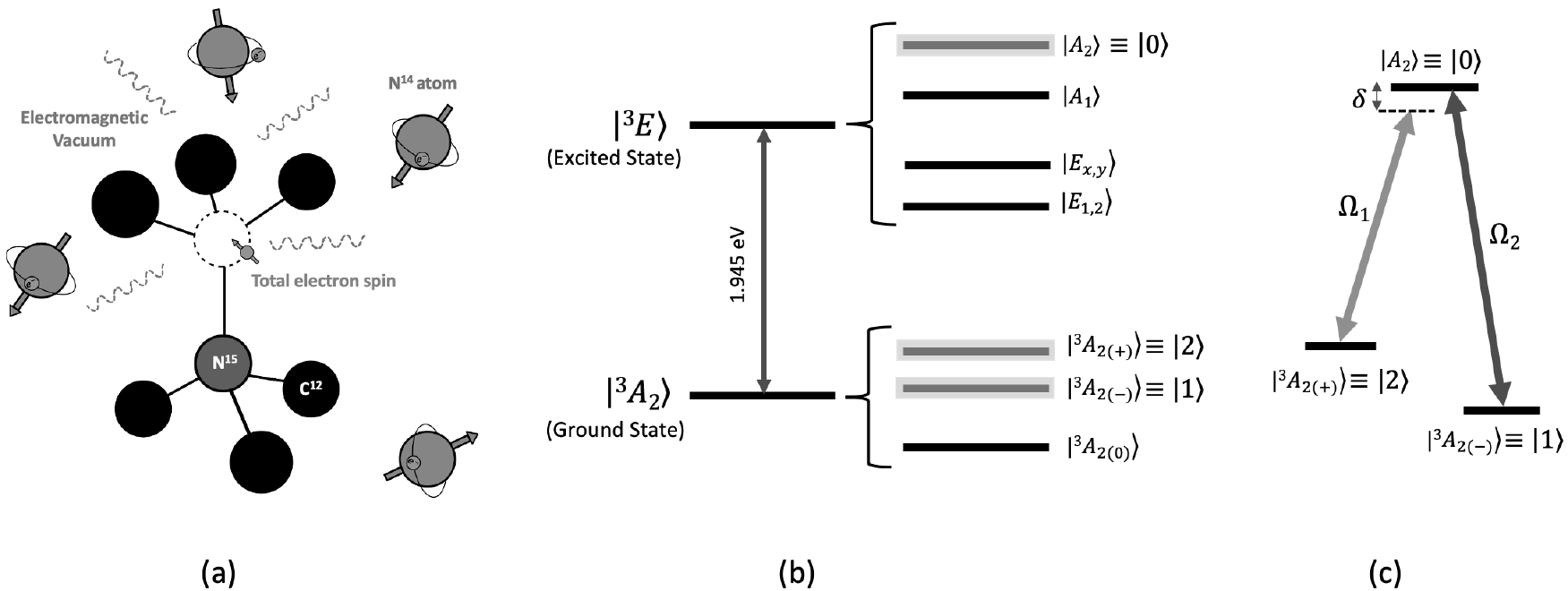}
    \vspace*{-3mm}
    \caption{(Color online) (a) Schematic of an NV center in the diamond lattice. Both the P1 centers electron spin bath and the quantum vacuum are explicitly shown. (b) Energy diagram of the ground and first excited states of the electronic structure of NV center (zero strain). (c) Energy levels of the selected $\Lambda$ system, driven by two coherent light sources with Rabi frequencies $\Omega_{1}$ and $\Omega_{2}$. $\delta$ denotes the two photon detuning.}
    \label{Figure1}
\end{figure*}

The NV center is embedded in the diamond lattice, the latter mostly composed of $^{12}$C isotopes (see Fig.~(\ref{Figure1}a)),
which have zero nuclear spins in their ground state. Thereby, the main sources of noise are due to paramagnetic impurities and nuclear spins in the lattice \cite{kehayias2015exploring}. The most common type of impurities are that of substitutional Nitrogen atoms (P1 centers) where, if present in concentrations of $>10$ppm, become the main source of noise. In contrast, if the diamond sample contains less Nitrogen impurities, then the main source of noise becomes the nuclear spins of the $^{13}$C isotopes ($I=1/2$), which makes about $1.1\%$ of the total Carbon atoms in the lattice \cite{maze2012free}. In what follows, we assume that the main source of noise is due to the P1 centers, each of which are comprised of a $^{14}$N nuclear spin ($I=1$) and an unpaired electron with spin $S=1/2$. Due to the fact that the gyromagnetic ratio of the electron is three orders of magnitude larger than that of nuclear spins, the electronic spins of P1 centers comprise the major spin bath producing the noise. In an ideal scenario, no photons are emitted when CPT is achieved. However, in the presence of noise, the expected value of the excited state population in CPT becomes larger than zero, and additional photons are emitted \cite{golter2013nuclear}.

Throughout our analysis, we assume the quantization axis to be along the NV center axis. It is well known that the NV center satisfies the $C_{3v}$ point-group symmetry \cite{ping2002group}; hence we shall use the corresponding group-theoretical notation. To briefly describe this point-group symmetry, we first note that the NV center remains unchanged if rotated around its central axis in multiples of $2\pi/3$ radians (see Fig.~(\ref{Figure1}a)). These rotations are usually denoted by $C_{3}$, $C_{3}^{2}$, and $C^{3}_{3}\equiv e$, corresponding to $2\pi/3$, $4\pi/3$, and $2\pi$, respectively. Here, $e$ is the identity operation, as a rotation by $2\pi$ radians is a trivial symmetry of the NV center. From Fig.~(\ref{Figure1}a), it is also clear that reflections with respect to each of the three planes passing through the NV-axis and any one of the neighboring Carbon atoms are also symmetry transformations of the NV center. These reflection operations are denoted by $\sigma_{1}$, $\sigma_{2}$, and $\sigma_{3}$. Therefore, the set of symmetry operations of the NV center is comprised of the collection of transformations $C_{3v}=\{e, C_{3}, C_{3}^{2}, \sigma_{1}, \sigma_{2}, \sigma_{3} \}$, which is easily verified to constitute a group. This symmetry group has been used in literature to analyze the electronic structure of the NV center, e.g. see \cite{maze2011properties}.

For the excited state of the $\Lambda$ system \cite{manson2006nitrogen, maze2011properties, chu2015quantum, doherty2011negatively}, we choose the spin-orbital state $A_{2}$ from the excited orbital manifold, motivated by the fact that it does not couple strongly to the non-radiative singlet states \cite{hincks2018statistical}. Combined with the two $m_{s}=\pm 1$ ground states $\{^{3\!}A_{2(-)}, ^{3\!}A_{2(+)}\}$ from the ground-state orbital manifold, this gives us a nearly perfectly closed $\Lambda$ system, which has already been demonstrated experimentally \cite{togan2010quantum, golter2013nuclear, golter2014optically}. We denote the states $ A_{2}$, $^{3\!}A_{2(-)}$, $^{3\!}A_{2(+)}$ by $|0\rangle, |1\rangle,$ and $|2\rangle$, respectively (see Fig.~ (\ref{Figure1}b)). We also make the tensor product between the orbital and spin degrees of freedom of the electronic structure explicit, by letting  
\begin{align}
|0\rangle &\equiv |A_{2}\rangle=|E_{-}\rangle \otimes |\!+1\rangle+|E_{+}\rangle \otimes |\!-1\rangle , \nonumber  \\
|1\rangle &\equiv|^{3\!}A_{2(+)}\rangle=|E_{0}\rangle \otimes |\!+1\rangle , \label{basis} \\
|2\rangle &\equiv|^{3\!}A_{2(-)}\rangle=|E_{0}\rangle \otimes |\!-1\rangle .
\nonumber 
\end{align}
Here, $|E_{0}\rangle$ and $|E_{\pm}\rangle$ are the orbital angular momentum eigenstates of the NV-center electron system \cite{maze2011properties}, labelled by eigenvalues $0$ and $\pm 1$ of $L^{(e)}_{z}$. The $|\!\pm 1\rangle$ states denote the two spin angular momentum eigenstates, labelled by the eigenvalues of $S^{(e)}_{z}$.

The Hamiltonian of the driven $\Lambda$ system is given by $H_{\Lambda}(t)\equiv H_{\Lambda}^0+H_{\text{drive}}(t)$ 
where, by assuming units $\hbar=1$, the two contributions take the form
\begin{eqnarray} 
H_{\Lambda}^0&=&\omega_{0}|0\rangle \langle 0|+\omega_{1}|1\rangle \langle 1|+\omega_{2}|2\rangle \langle 2| , \notag\\
        H_{\text{drive}}(t)&= & ({\Omega_{1}}/{2})\, e^{-i(\omega_{L1}t+\phi_{1})}|1\rangle \langle 0| \label{eq:Ham0} \\
        &+&  ({\Omega_{2}}/{2})\, e^{-i(\omega_{L2}t+\phi_{2})}|2 \rangle \langle 0| + \text{H.c.}, \notag
\end{eqnarray}
where $\Omega_{1}$ and $\Omega_{2}$ are the Rabi frequencies, and ($\omega_{L1}, \phi_{1}$) and ($\omega_{L2}, \phi_{2}$) the frequencies and phases of the two coherent light sources, respectively. Furthermore, we denote by $\delta_{1(2)}=\omega_{L1(2)}-\omega_{1(2)}$ the detuning of the first (second) laser. As mentioned in the introduction, we shall work under the assumption that any control errors arising in the implementation of $H_{\text{drive}}(t)$ may be neglected in comparison with environmental noise. To justify this assumption, we recall that control errors in implementing $H_{\text{drive}}(t)$ have been discussed previously in literature. In \cite{dalton1982effects}, the authors showed that, in the presence of stochastic control errors in the frequencies $\omega_{L1}$ and $\omega_{L2}$ in $H_{\text{drive}}(t)$, the CPT will be destroyed if the two laser field fluctuations are uncorrelated. However, they also showed that CPT can be restored if a certain amount of critical cross-correlation between the two laser fields is achieved experimentally. This has been shown to be possible if the two laser fields are derived from a shared source via well-known accousto-optic modulation techniques, which has been achieved in e.g. \cite{thomas1982observation}. In our article, we aim at analyzing the possibility of accomplishing perfect CPT in principle, even when such experimental techniques have been implemented. This means finding out the contribution of the implementation medium (i.e. the spin bath) in the imperfection of the observed CPT. Furthermore, we also note that in \cite{dalton1985coherent}, the authors showed that even for unequal laser bandwidths, the CPT will still be preserved. Hence, we neglect these types of control errors for the purposes of our analysis, provided that we are exclusively interested in the CPT phenomenon of a $\Lambda$ system.

We model the effect of the P1 center electron spin bath \cite{wang2012comparison, dobrovitski2008decoherence, wang2013spin, hanson2008coherent, simanovskaia2013sidebands} on the $\Lambda$ system as a fluctuating classical magnetic field. 
The value of this field at the position of the NV center at time $t$ is denoted by $b(t)$. Specifically, we assume the spin-bath noise to be zero-mean, stationary, and sufficiently weak to be treated perturbatively (see Sec.~\ref{sub:tcl}). In particular, the lowest-order (two-point) correlation function 
is determined by 
\begin{equation}
C(t_1,t_2)\equiv {\mathbb E}\{b(t_1)b(t_2)\} = C(|t_1-t_2|), 
\label{eq:cor}
\end{equation}
where ${\mathbb E}$ denotes the ensemble average over realizations of the classical stochastic process $\{b(t)\}$. In the presence of this stochastic bath, the Hamiltonian of the driven $\Lambda$ system is then given by $H(t)\equiv H_{\Lambda}(t)+H_{c}(t)$, where 
\begin{equation}
H_{c}(t)\equiv H_c[b(t)]= -\mu_{B}(L_{z}+2S_{z})\, b(t), 
\label{eq:Hn}
\end{equation}
is the semi-classical interaction Hamiltonian describing the coupling of the electronic system to the field,  
with $L_{z}$, $S_{z}$, and $\mu_{B}=\frac{e\hbar}{2m_{e}c}$ being the angular momentum, spin operators, and the Bohr magneton, respectively. We write such a Hamiltonian in the $\Lambda$-system basis of Eqs. \eqref{basis} as
\begin{equation}
   H_{c}(t)=\sum_{i,j=0,1,2}\!\langle i|H_{c}(t)|j\rangle  |i\rangle \langle j|  . 
\end{equation}
Using the fact that the expectation value of the operators $L^{(e)}_{z}$ and $S^{(e)}_{z}$ are given by $(0, 0), (0, 1),$ and $(0, -1)$ for the states $|0\rangle, |1\rangle,$ and $|2\rangle$, respectively, along with 
the orthonormality of the states $|E_{0}\rangle$ and $|E_{\pm}\rangle$, we arrive at
\begin{equation}
H_{c}(t)=-\gamma_{e}b(t)\,|1\rangle \langle 1|+\gamma_{e}b(t)\,|2\rangle \langle 2| , 
\end{equation}
where $\gamma_{e}=\frac{e\hbar}{m_{e}c}$ is the gyromagnetic ratio of the electron. Physically, $\gamma_{e}b(t)$ is the time-dependent frequency fluctuation of the $\Lambda$-system ground states; see  Fig.~(\ref{Figure1}b).


\section{Noisy Coherent Population Trapping}
\label{NCPT}

\subsection{Master equation for ideal CPT dynamics}

As mentioned, CPT is an equilibration phenomenon in a driven three-level $\Lambda$ system where, irrespective of the initial state \cite{jyotsna1995coherent,ticozzi2012hamiltonian}, the dynamics gets restricted to the two-ground-state manifold. Physically, this is a quantum-mechanical consequence of the destructive interference between the two transition probability amplitudes from individual ground states to the same excited state in the $\Lambda$ system. 
In order to set the stage for the noisy setting, we briefly review the derivation of a quantitative model within a ME formalism.  

In the presence of spontaneous decay alone, the system and the bath are described by the total Hamiltonian 
\begin{equation}
H^0_{\text{tot}}(t) \equiv H_\Lambda^0+H_{\text{drive}}(t)+H_{\text{vac}}+H_{\Lambda\text{-vac}},
\label{Htot0}
\end{equation}
where $H_{\text{vac}}$ is the Hamiltonian of the electromagnetic vacuum and $H_{\Lambda\text{-vac}}$ is the interaction Hamiltonian between the $\Lambda$ system and the vacuum, in the standard dipole approximation. We write the ME in the interaction picture with respect to $H_{\Lambda}^0+H_{\text{vac}}$ and denote an operator $X$ in this representation by $\tilde{X}$. The Liouville-von Neumann equation describing the evolution of the driven $\Lambda$ system and the vacuum is then given by
\begin{equation} 
\dot{\tilde{\sigma}}(t)=\hat{L}^{0}\tilde{\sigma}(t)=-i[\tilde{H}_{\text{drive}}(t)+\tilde{H}_{\Lambda\text{-vac}}(t), \tilde{\sigma}(t)] , \label{eqn:liouville-von}
\end{equation}
where $\tilde{\sigma}(t)$ is the density matrix of the total system and $\hat{L}^{0}$ is the Liouvillian superoperator corresponding to the Hamiltonian $\tilde{H}^{0}_{\text{tot}}$ in the interaction picture
\begin{equation}
    \hat{L}^{0}(\bullet)=-i[\tilde{H}_{\text{drive}}(t)+\tilde{H}_{\Lambda\text{-vac}}(t), (\bullet)].
\end{equation} 

By assuming that the joint initial state $\sigma(0)=\rho(0)\otimes \rho_{\text{vac}}$ is product, 
and treating the coupling to the vacuum in the standard Born-Markov approximation \cite{breuer2002theory}, the resulting reduced dynamics is given by a Lindblad ME of the form 
\begin{equation}
    \frac{d}{dt}{\text{Tr}}_{\text{vac}}\tilde{\sigma}(t)=\dot{\tilde{\rho}}(t)=-i[\tilde{H}_{\text{drive}}(t), \tilde{\rho}(t)]+R_q[\tilde{\rho}(t)] ,
    \label{eqn:A1}
\end{equation}
where the Hamiltonian may be explicitly computed from Eq. \eqref{eq:Ham0} and 
\begin{equation}
R_q[\tilde{\rho}]\equiv -\frac{1}{2}\sum _{i=1,2}
\,(L^{\dagger}_{i}L_{i}\tilde{\rho}+\tilde{\rho}L^{\dagger}_{i}L_{i}-2L_{i}\tilde{\rho}L^{\dagger}_{i})  
\label{eq:Rq}
\end{equation}
is the dissipator accounting for the quantum Markovian environment. The two Lindblad operators are given by
\begin{equation}
L_{i}=\sqrt{{\Gamma}/{2}} \, |i\rangle\langle 0| , \quad \text{for} \hspace{0.5cm} i=\left\{1,2 \right\}, 
\label{eq:lind}
\end{equation}
where $\Gamma=\Gamma_{01}\approx \Gamma_{02}$ is the decay rate from the excited state to each of the ground states (explicitly,  $\Gamma_{0i}=(\omega_{0}-\omega_{i})^{3}d^{2}_{0i}/3\pi \varepsilon_{0}\hbar c^{3}$, $i=1,2$, where $d_{0i}$ is a matrix element from the dipole coupling matrix \cite{breuer2002theory}). By dropping, for simplicity, the tilde notation for interaction-picture operators, the result is a coupled set of differential equations for the density matrix elements. In particular, in the relevant case where the detunings of the two lasers are $\delta_{1}=0$ and $\delta_{2}\equiv \delta$, and $\phi_1 = \phi_2 =0$, we recover the known expressions (see, for instance, \cite{arimondo1996v, brewer1975coherent}):
\begin{align}
    \dot{\rho}_{00} &=-\Gamma \rho_{00}+i{\Omega_{1}}/{2}\rho_{01}+i{\Omega_{2}}/{2}\rho_{02}+c.c. \label{eqn:system1},\\
\dot{\rho}_{11} &={\Gamma}/{2} \rho_{00}-i{\Omega_{1}}/{2}\rho_{01}+c.c.\label{eqn:system2},\\
 \dot{\rho}_{22} &={\Gamma}/{2} \rho_{00}-i{\Omega_{2}}/{2}\rho_{02}+c.c.\label{eqn:system3},\\
 \dot{\rho}_{01} &=-{\Gamma}/{2} \rho_{01}+i{\Omega_{1}}/{2}(\rho_{00}-\rho_{11})-i{\Omega_{2}}/{2}\rho_{21}\label{eqn:system4},\\
  \dot{\rho}_{02} &=-({\Gamma}/{2}+i\delta) \rho_{02}+i{\Omega_{2}}/{2}(\rho_{00}-\rho_{22})-i{\Omega_{1}}/{2}\rho_{12}\label{eqn:system5},\\
   \dot{\rho}_{12} &=i\delta \rho_{12}-i{\Omega_{1}}/{2}\rho_{02}+i{\Omega_{2}}/{2}\rho_{10}.\label{eqn:system6}
\end{align}
The above set of coupled differential equations can be compactly represented in matrix form as  
\begin{equation}
\dot{\vec{\rho}}=\hat{A}\vec{\rho},\qquad \vec{\rho}\equiv(\rho_{00}, \rho_{01}, \ldots, \rho_{22}),
\label{eq:rho}
\end{equation}
in terms of the vectorized density matrix.

It is well known that for a Lindblad ME as in Eq.~ \eqref{eqn:A1}, a steady-state solution always exists, and it is globally attractive if and only if it is unique \cite{Sophie}. Numerically, we have explicitly 
verified that $\det{\hat{A}}=0$ for arbitrary values of the parameters $\Gamma, \Omega_{1}, \Omega_{2}$, and $\delta$. In particular, this implies that the experimentally tunable parameter $\delta$ can be varied freely and there will be a steady state $\vec{\rho}^{\,\text{eq}}_{\delta}$, determined by $\hat{A}\vec{\rho}^{\,\text{eq}}_{\delta}=0$. 
Unsurprisingly, the steady state corresponding to $\delta=0$ is the dark state 
\begin{equation}
    |d \rangle\equiv\frac{\Omega_{2}}{\Omega}|1\rangle-\frac{\Omega_{1}}{\Omega} |2\rangle, \quad  \Omega\equiv \sqrt{\Omega^{2}_{1}+\Omega^{2}_{2}},
\end{equation}
because the CPT condition of having a zero two-photon detuning $\delta_{12}=\delta_{1}-\delta_{2}$ is equivalent to having $\delta=0$. By invoking the sufficient conditions for uniqueness provided in \cite{Sophie}, one may verify that the Lindblad dynamics has the unique steady state $\vec{\rho}^{\,\text{eq}}_{0}=|d\rangle\langle d|$ independent of $\Gamma$, provided that $\Gamma \ne 0$, in which case this steady state is reached from an arbitrary initial preparation. 

\subsection{Master equation for noisy CPT dynamics}
\label{sub:tcl}

Stochastic bath models have been extensively discussed in the literature \cite{kubo1963stochastic, 
van1992stochastic, saeki1988stochastic, breuer2002theory, dalton1982effects, mishra2014three}.  
In the present setting, to integrate the coupling to a quantum (vacuum) environment and to the classical spin-bath environment into a single equation for the reduced dynamics of the $\Lambda$ system, 
we start from the full stochastic Hamiltonian  
\begin{equation}
H_{\text{tot}}(t)= H_{\text{tot}}^0(t) + H_{c}[b(t)], 
\end{equation}
where the noiseless Hamiltonian and the noise term are given by Eqs.~\eqref{Htot0} and \eqref{eq:Hn}, respectively. Moving to the interaction picture with respect to the total free Hamiltonian $H^0_{\Lambda}+H_{\text{vac}}$, and denoting the full density matrix for a \textit{single realization} of the stochastic process $\left\{b(s)\right\}_{s=0}^{s=t}$ by $\sigma(t;\left\{b(t)\right\})$, the formal solution of the Liouville-von Neumann equation reads
\begin{equation}
    \sigma(t;\left\{b(t)\right\})=\mathcal{T}\exp{\left\{ \int_{0}^{t}ds \hat{L}(s) \right\}}\sigma(0),
\end{equation}
where $\mathcal{T}$ denotes time ordering and $\hat{L}$ is the Liouvillian superoperator corresponding to the interaction part of $H_{\text{tot}}(t)$ in the interaction picture 
\begin{equation}
\begin{aligned}
    \hat{L}(\bullet)&=-i[\tilde{H}_{\text{drive}}(t)+\tilde{H}_{\Lambda\text{-vac}}(t)+\tilde{H}_{c}[b(t)], (\bullet)],\\
    &=\hat{L}^{0}(\bullet)-i[\tilde{H}_{c}[b(t)], (\bullet)]. \label{eqn:liouvillian}
\end{aligned}
\end{equation}
Let us define the projection superoperator $\hat{P}$ by requiring that 
$\hat{P}\sigma\equiv ({\text{Tr}}_{B}\sigma) \otimes \rho_{B}$, for arbitrary $\sigma$ and fixed $\rho_B$ 
(the latter is usually taken to be the stationary Gibbs state of the quantum bath). Assuming as before that  $\sigma(0)=\rho(0)\otimes \rho_{B}$, we get $\hat{P}\sigma(0)=\sigma(0)$. From here on, we shall use van Kampen's cumulant notation (that is, $\langle \hat{\chi} \rangle =\hat{P}\hat{\chi}\hat{P}$). We apply the projection operator to both sides of the previous equation and use the property $\hat{P}^{2}=\hat{P}$ to get
\begin{equation}
    \rho(t;\left\{b(t)\right\})\otimes\rho_{B}=\langle \mathcal{T}\exp{\left\{ \int_{0}^{t}ds \hat{L}(s) \right\}}\rangle \hat{P}\sigma(0). \label{eqn:don't_diff}
\end{equation}

Next, we average both sides of this equation with respect to the classical noise first 
and then follow steps that are well known, due to van Kampen \cite{van1974cumulant}. Starting from the ensemble-averaged equation
\begin{equation}
{\mathbb E}(\hat{P}\sigma(t)) = {\mathbb E}\left\{ \langle {\cal T}\exp\left\{ \int_{0}^{t}ds \hat{L}(s) \right\}\rangle \right\}  \hat{P}\sigma(0), \label{eqn:averaged_liouville}
\end{equation}
we expand the right hand-side to 
\begin{equation}
(I+E_{0}(\hat{1})+E_{0}(\hat{1}, \hat{2})+\ldots)\hat{P}\sigma(0), \label{eqn:ensemble_eqn}
\end{equation}
where $ E_{0}(\hat{1}, \hat{2}, \ldots, \hat{n})$ is the $n$-th term in the Dyson expansion, given by 
\begin{equation}
\int^{t}_{0}dt_{n}\int^{t_{n}}_{0}dt_{n-1}\ldots\int^{t_{2}}_{0}dt_{1}{\mathbb E} \left\{ \langle \hat{L}(t_{n})\ldots\hat{L}(t_{2})\hat{L}(t_{1}) \rangle \right\}.
\end{equation}
On the other hand, differentiating Eq.~\eqref{eqn:averaged_liouville} yields 
\begin{equation}
\frac{d}{dt}{\mathbb E}(\hat{P}\sigma(t;\left\{b(t)\right\})) =
 (E_{1}(\hat{1})+E_{1}(\hat{1}, \hat{2})+\ldots)\hat{P}\sigma(0), 
\label{eqn:derivative_of_ensemble_eqn}
\end{equation}
where $ E_{1}(\hat{1}, \hat{2}, \ldots, \hat{n})$ denotes the time derivative of $ E_{0}(\hat{1}, \hat{2}, \ldots, \hat{n})$. 
We can rewrite the right hand-side of Eq.~(\ref{eqn:derivative_of_ensemble_eqn}) in terms of ${\mathbb E}(\hat{P}\sigma(t)) \equiv 
\sigma_{\text{av}}(t)$ by solving Eq.~(\ref{eqn:averaged_liouville}) with respect to $\hat{P}\sigma(0)$ as
\begin{equation}
    \hat{P}\sigma(0)=\left\{1+E_{0}(\hat{1})+E_{0}(\hat{1},\hat{2})+\ldots \right\}^{-1}\sigma_{\text{av}}(t).
\end{equation}
Upon substituting the result back into Eq.~(\ref{eqn:derivative_of_ensemble_eqn}), we obtain the TCL ME, 
\begin{equation}
\dot{\sigma}_{\text{av}}(t) =\hat{\kappa}(t) \sigma_{\text{av}}(t).
\end{equation}
The TCL generator $\hat{\kappa}(t)$ is determined in terms of the (van Kampen) cumulants of the Liouvillian superoperator as 
\begin{equation}
    \left\{E_{1}(\hat{1})+E_{1}(\hat{1}, \hat{2})+\ldots\right\} \left\{1+E_{0}(\hat{1})+E_{0}(\hat{1},\hat{2})+\ldots \right\}^{-1}, \label{eqn:expansion}
\end{equation}
and can be expanded in orders of the interaction coefficients, $\hat{\kappa}(t)=\sum_{n}\hat{\kappa}_{n}(t)$. A common way of writing this expansion is given by the ordered cumulants $\langle \langle \hat{L}(t) \hat{L}(t_{n-1})\ldots\hat{L}(t_{2})\hat{L}(t_{1}) \rangle \rangle_{\text{oc}}$ defined implicitly by equating $\hat{\kappa}_{n}(t)$ to
\begin{equation}
    \int^{t}_{0}dt_{n-1}\ldots \int^{t_{2}}_{0}dt_{1}\mathbb{E}\left\{\langle \langle \hat{L}(t) \hat{L}(t_{n-1})\ldots\hat{L}(t_{1})\rangle \rangle_{\text{oc}}\right\},
\end{equation}
where an explicit formula of ordered cumulants is found in \cite{breuer2002theory}. The first three terms of the expansion $\hat{\kappa}(t)=\sum_{n}\hat{\kappa}_{n}(t)$ following Eq.~(\ref{eqn:expansion}) are given by
\begin{align}
    \hat{\kappa}_{1}(t)&=E_{1}(\hat{1}),\\
    \hat{\kappa}_{2}(t)&=E_{1}(\hat{1},\hat{2})-E_{1}(\hat{1})E_{0}(\hat{1}),
\end{align}
\vspace{-8mm}
\begin{equation}
\begin{aligned}
\hat{\kappa}_{3}(t)=&E_{1}(\hat{1}, \hat{2}, \hat{3})-E_{1}(\hat{1}) E_{0}(\hat{1}, \hat{2})\\
&-E_{1}(\hat{1}, \hat{2}) E_{0}(\hat{1})+E_{1}(\hat{1}) E_{0}^{2}(\hat{1})
\end{aligned}
\end{equation}

We emphasize that $\hat{\kappa}(t)$ depends on \emph{both} the stochastic process $b(t)$ and the coupling coefficient to the vacuum field $\Gamma$. We make the assumption $\mathrm{Tr}_{B}\rho_{B}\tilde{H}_{\Lambda-\text{vac}}(t)=0$ \cite{breuer2002theory} and use it, along with the definition of $\hat{P}$, Eq.~(\ref{eqn:liouvillian}), and $\mathbb{E}\left\{b(t)\right\}=0$, to compute $E_{1}(\hat{1})\sigma_{\text{av}}(t)$ as
\begin{equation}
    \mathbb{E}\left\{\hat{P}\hat{L}(t)\hat{P}\right\}\sigma_{\text{av}}(t)=-i[\tilde{H}_{\text{drive}}(t), \mathrm{Tr}_{B}\sigma_{\text{av}}]\otimes \rho_{B}. 
\end{equation}
From the above, we can easily show
\begin{equation}
    E_{0}(\hat{1})\sigma_{\text{av}}(t)=\int_{0}^{t}ds\mathbb{E}\left\{\hat{P}\hat{L}(s)\hat{P}\right\}\sigma_{\text{av}}(t)=0,
\end{equation}
due to $\int^{t}_{0}ds\exp\left\{\pm i(\omega_{L1(2)}+\omega_{0}-\omega_{1(2)})s\right\}=0$ for $t>>1/\omega_{L1(2)}$ since $\omega_{L1(2)}$ is in the optical range,
which leads to $\hat{\kappa}_{2}(t)=E_{1}(\hat{1},\hat{2})$, etc.
For sufficiently \emph{weak coupling} to both the classical and quantum baths,
we truncate the TCL generator expansion at the second order. To justify this, let us introduce the correlation time $\tau_{c}$ by assuming that $\mathbb{E}\left\{ \langle \langle \hat{L}(t)\hat{L}(t_{n-1})\ldots\hat{L}(t_{1}) \rangle \rangle_{\text{oc}} \right\}$ vanishes whenever the difference between any two of the times $t_{1}, t_{2}, \ldots, t_{n-1}$ is larger than $\tau_{c}$. Note that if the quantum and classical baths have correlation times $\tau^{cl}$ and $\tau^{q}$ respectively, then the correlation time $\tau_{c}=\max\left\{ \tau^{cl}, \tau^{q}\right\}$. Each $\mathbb{E}\left\{\langle \langle \hat{L}(t)\hat{L}(t_{n-1})\ldots\hat{L}(t_{1}) \rangle \rangle_{\text{oc}} \right\}$ is of the order of the coupling strength $g$ to the power $n$ (where $g=\max \left\{g^{cl}, g^{q} \right\}$ with $g^{cl}$ and $g^{q}$ being the individual couplings of the classical and quantum baths to the $\Lambda$ system). This implies that the $n$-th term $\hat{\kappa}_{n}(t)$ is of the order of $g^{n}\tau_{c}^{n-1}$ and hence $\hat{\kappa}(t)=\sum_{n}\hat{\kappa}_{n}(t)$ is an expansion with respect to the dimensionless quantity $g\tau_{c}$ \cite{van1974cumulant} ($g\tau_{c}<<1$ implies the existence of the two-time scale $\tau_{c}<<\frac{1}{g}= \tau_{\Lambda}$ where $\tau_{\Lambda}$ is the characteristic time of the $\Lambda$ system dynamics, due to the coupling with both baths). Ignoring $\hat{\kappa}_{3}(t)$ with respect to $\hat{\kappa}_{2}(t)$ is possible if $\hat{\kappa}_{3}/\hat{\kappa}_{2} \thicksim g\tau_{c}<<1$, which can be interpreted (equivalently) either as a short correlation time for a fixed coupling $\tau_{c}<<\frac{1}{g}$ or a weak coupling for a fixed correlation time $g<<\frac{1}{\tau_{c}}$. The former interpretation is usually picked \cite{kryszewski2008master} because this is later used to complete the Born-Markov approximation by taking the upper limit of integration in the resulting integro-differential equation to infinity. However, to arrive at our main equation, we make no such approximation. For 
    $\hat{\kappa}_{2}(t) = E_{1}(\hat{1}, \hat{2})$
we have
\begin{equation}
    E_{1}(\hat{1}, \hat{2})\sigma_{\text{av}}(t)= \int_{0}^{t} \!ds \, {\mathbb E} \{ \hat{P}\hat{L}(t)\hat{L}(s)\hat{P} \}\sigma_{\text{av}}(t) .
\end{equation}
By tracing over the quantum bath and using $\mathbb{E}\left\{b(t) \right\}=0$, we obtain 
\begin{equation}
    -\int_{0}^{t}\!ds \,{\text{Tr}}_{B} {\mathbb E}\left\{[\tilde{H}_{\text{int}}(t), [\tilde{H}_{\text{int}}(s), \rho_{\text{av}}(t)\otimes \rho_{B}]]\right\},
\end{equation}
\vspace{-8mm}
\begin{equation}
\begin{aligned}
    &=-\int_{0}^{t}\!ds \,{\text{Tr}}_{B} {\mathbb E}\left\{[\tilde{H}^{0}_{\text{int}}(t), [\tilde{H}^{0}_{\text{int}}(s), \rho_{\text{av}}(t)\otimes \rho_{B}]]\right\} \\
    &+\int_{0}^{t}\!ds \,{\text{Tr}}_{B} {\mathbb E}\left\{[\tilde{H}_{c}[b(t)], [\tilde{H}_{c}[b(s)], \rho_{\text{av}}(t)\otimes \rho_{B}]]\right\},
\end{aligned}
\end{equation}
where $\rho_{\text{av}}(t)\equiv {\text{Tr}}_{B}\sigma_{\text{av}}(t)$ is the reduced density matrix of the $\Lambda$ system and \( \tilde{H}_{\text{int}}(t)=\tilde{H}^{0}_{\text{int}}(t)+\tilde{H}_{c}[b(t)]=\tilde{H}_{\text{drive}}(t)+\tilde{H}_{\Lambda\text{-vac}}(t)+\tilde{H}_{c}[b(t)] \). Hereafter, we omit the subscript $``\text{av}"$ for notational convenience. Noting that
\begin{equation}
    \tilde{H}_{c}[b(t)]=e^{ iH^0_{\Lambda}t}H_{c}[b(t)]e^{ -iH^0_{\Lambda}t}=H_{c}[b(t)]=\gamma_{e}b(t)Z_{12} ,
\end{equation}
and $Z_{12}\equiv |1\rangle \langle 1|-|2\rangle \langle 2|$. We thus easily arrive at
\begin{equation}
    \dot{\rho}(t)=-i[H_{\text{drive}}(t), \rho(t)]+R_{q}[\rho(t)]+R_{c}[\rho(t)], 
    \label{eqn:rate}
\end{equation}
where $R_q$ is the Lindblad dissipator already specified in Eqs.~\eqref{eq:Rq}-\eqref{eq:lind}, whereas 
\begin{align}
R_{c}[\rho (t)] & = -\gamma^{2}_{e}\alpha(t) \left[ Z_{12}^2 \rho (t)+\rho(t) Z_{12}^2 - 2 Z_{12} \rho (t) Z_{12} \right],
\end{align}
is the time-local dissipator accounting for the additional spin-bath noise. Here, the time-dependent strength parameter is given by 
\begin{equation}
\alpha(t) \equiv \int_{0}^{t}\!\! ds\,C(|t-s|) = 2\int_0^\infty \!d\omega \, S(\omega) \frac{\sin \omega t}{\omega}, 
\label{eqn:alpha}
\end{equation}
in terms of the noise correlation function $C(t)$ of Eq.~ \eqref{eq:cor} and the corresponding noise spectral density, determined by the Fourier transform 
\begin{equation}
    S(\omega)=\frac{1}{2\pi}\int_{-\infty}^{+\infty} \!\! d\tau \, C(|\tau|) e^{-i\omega \tau}.
\end{equation}

Physically, the noise parameter $\alpha(t)$ in the ME is the only term that carries information about the history of the stochastic magnetic field, consistent with the fact that no Markovian assumption is involved in the TCL ME.  
%
\begin{figure}[ht!]
    \centering
    \includegraphics[width=8.0cm]{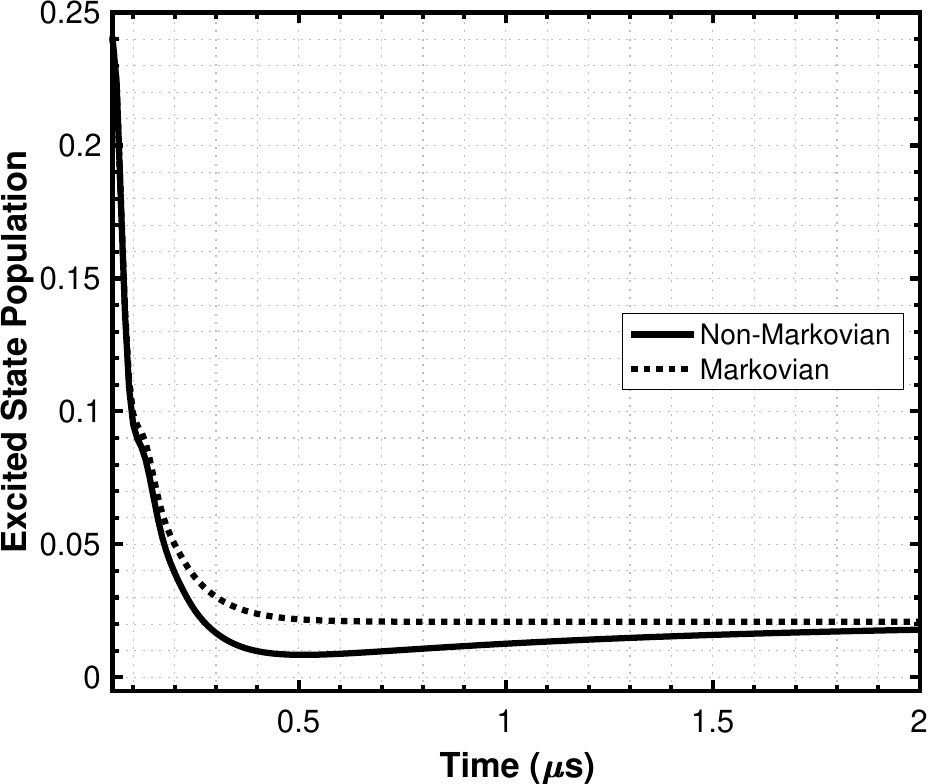}
    \caption{(Color online) Excited-state population as a function of time. The bold line represents the solution of Eq.~(\ref{eqn:rate}), the dotted line above it is the solution of the same equation in the Born-Markov approximation when $\alpha(t) \approx \alpha$. Note the important differences in the equilibration dynamics (slower equilibration) when non-Markovian effects are introduced. Parameter values are as follows: $\Omega_{1}=\Omega_{2}=46$MHz, $\Gamma /2\pi=7$MHz, and $\delta=0$.}
    \label{fig:dynamics}
\end{figure}
Similar to Eq.~\eqref{eq:rho}, Eq.~\eqref{eqn:rate} can still be cast as a linear system of coupled differential equations,
\begin{equation} 
\dot{\vec{\rho}}=\hat{A^{\prime}}(t)\vec{\rho}, \qquad \vec{\rho}\equiv(\rho_{00}, \rho_{01}, \ldots, \rho_{22}),
\label{ME2}
\end{equation}
in terms of a new superoperator matrix $\hat{A^{\prime}}(t)$. However, the dynamical system is now \emph{time-varying} in general, due to the time dependence encoded in $\alpha(t)$, which in turn stems from the colored spectrum. Characterizing the steady states and their stability becomes a significantly less straightforward problem \cite{LTV}, which is beyond our present scope. 
As an illustration of the influence that bath properties may have on the transient dynamics, we showcase in Fig.~(\ref{fig:dynamics}) the dynamics of the excited state population obtained by solving Eq.~\eqref{ME2} for an exponentially decaying correlation function, $C(t)=c_0^{2}\exp{(-t/\tau_{c})}$, 
with $\gamma_{e}c_0\approx 0.5$MHz and $\tau_{c}=\tau^{cl}\approx 1 \mu$s \cite{simanovskaia2013sidebands, wang2013spin, dobrovitski2008decoherence, wang2012comparison}. 
The comparison is done between the solution of Eq.~(\ref{eqn:rate}) with and without making the Born-Markov approximation $\alpha(t) \approx \alpha$. For the given correlation function $\alpha(t)=c^{2}_{0}\tau_{c}[1-\exp{(-t/\tau_{c})}]$, whereas $\alpha=c^{2}_{0}\tau_{c}$. Our simulation shows that the excited state population exhibits a slower approach to equilibrium in the non-Markovian regime (for the given experimental parameters), hence a lower photon count overall. Such a non-Markovian equilibration effect might be observed by real-time photon count experiments such as in \cite{lekavicius2017transfer}. 

Since our main focus is CPT, which is an equilibrium phenomenon, the steady state will be seen in the long-time (effectively Markovian) limit, whereby
\begin{equation}
    \alpha(t)= \int^{t}_{0}\!\!d\tau \,C(\tau) \approx 
\int^{\infty}_{0}\!\!d\tau \,C(\tau)\equiv \alpha  = S(0)  , \quad t \rightarrow \tau_c.
\end{equation}
This integral is often encountered when calculating the decoherence time for a two-level system in the presence of Gaussian dephasing \cite{breuer2002theory}, as $T^{-1}_{2}=\gamma^{2}_{e}\alpha$. 
More generally, $\alpha$ may be related to the fastest decoherence timescale of an arbitrary N-level system, which is the timescale over which $\rho (t)$ changes appreciably due to the coupling to the bath \cite{mozgunov2020completely}, denoted by $\tau_{2}\equiv (\gamma^{2}_{e}\alpha)^{-1}$ for our three-level system. In our case, this parameter is treated like a noise parameter rather than the characteristic decay time of the $\Lambda$ system, because the spin bath is not the only bath acting on it.

\section{Applications}

In this section, we illustrate how the theoretical description of CPT dynamics developed so far may be applied to two applications of independent interest. 

\subsection{Parametric noise estimation}
\label{applications1}

\begin{figure}[t]
    \centering
    \includegraphics[width=8.6cm]{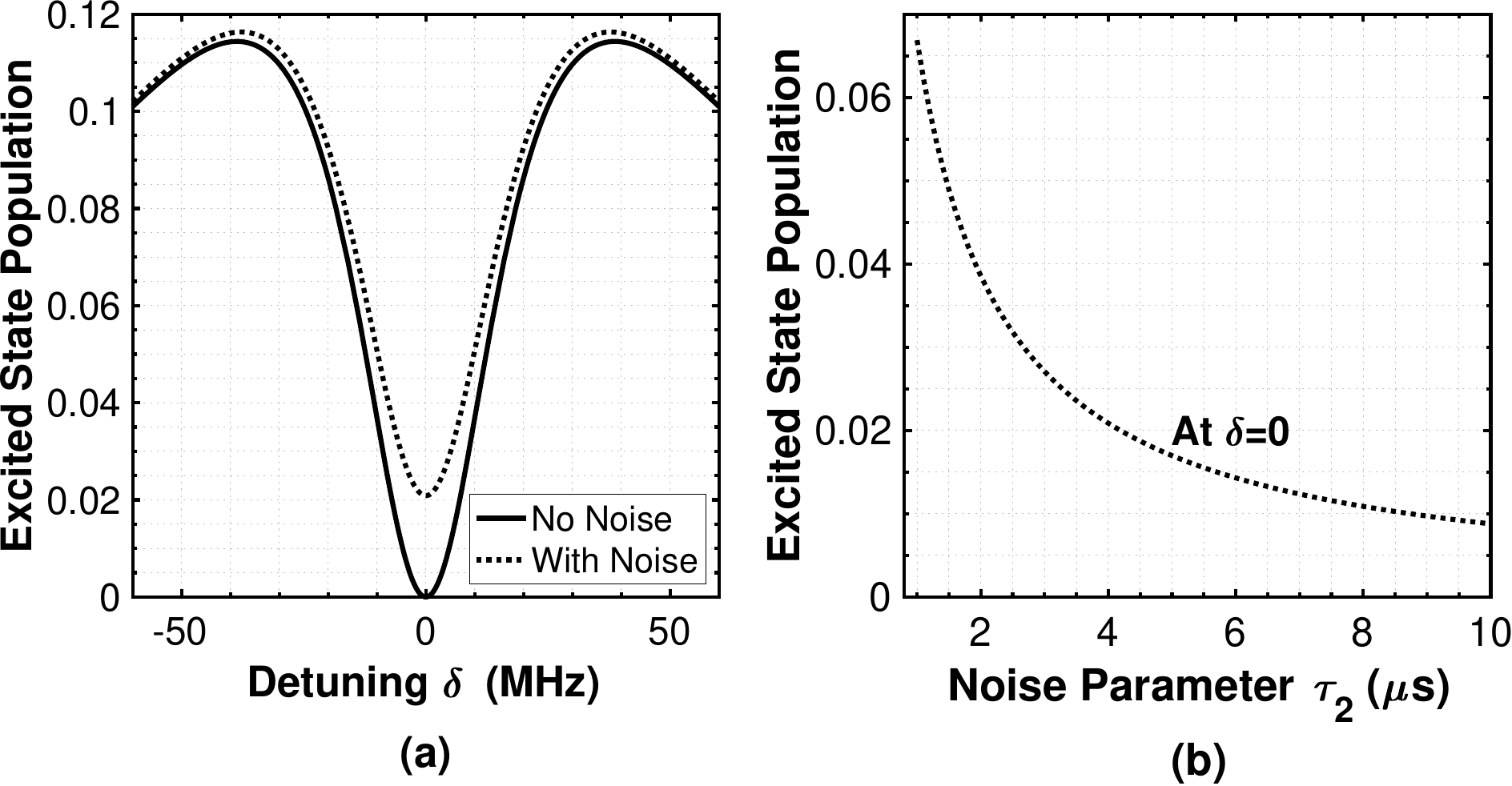}
    \vspace*{-3mm}
    \caption{(Color online) (a) Excited-state population of the $\Lambda$ system versus the two-photon detuning of the two lasers; the second laser detuning is taken to be $\delta_{2}=0$. The depth of the CPT dip depends on the value of the noise parameter characterizing the classical bath. (b) Excited-state population as a function of $\tau_{2}={1}/({\gamma^{2}_{e}\alpha})$ at the CPT dip ($\delta=0$). The values for the experimental parameters are motivated by \cite{golter2014optically, childress2006coherent} and taken to be $\Gamma/2\pi=7\,$MHz and $\Omega_{1}=\Omega_{2}=46\,$MHz.} 
    \label{figure2}
\end{figure}

First, by determining the steady-state solution of the ME Eq.~(\ref{eqn:rate}), the equilibrium excited-state population may be studied as a function of relevant parameters, in particular, the detuning. In Fig.~(\ref{figure2}a), representative results are shown for CPT dynamics with and without the presence of the spin-bath noise. Notably, in the noisy case, the excited-state population no longer vanishes; rather, the characteristic CPT dip has a finite height from zero. The depth of this dip depends on the value of the noise parameter $\alpha$. This is caused by the fluctuating magnetic field randomly shifting the two ground states of the $\Lambda$ system, and spoiling the destructive interference condition necessary for CPT. Interestingly, a similar steady-state behavior of driven $\Lambda$ systems was reported in \cite{blaauboer1997steady} in the presence of an incoherent optical pumping between only one of the ground states and the excited state. Likewise, 
including the decoherence of the ground states would also lead to a similar effect \cite{xu2008coherent}. These effects will play a negligible role (if any) in the CPT setting we consider. 
On the one hand, no incoherent optical pumping is present in our scheme.
On the other hand, the effect reported in \cite{xu2008coherent} 
is negligible due to the relatively fast equilibration time of the $\Lambda$ system in the NV center when compared to $T^{\star}_{2}$ times  \cite{wang2013spin}. It is also worth noting that Ref.\cite{dalton1982effects} arrives at a similar effect on CPT, taking the noise source to be the fluctuations of the laser frequencies instead. Furthermore, they show that some critical cross-correlations between the two lasers is necessary to remove this effect experimentally \cite{kennedy1984cross}. In contrast, the effect shown in this article is intrinsic to the $\Lambda$ system implementation medium, e.g. the (random) distribution of P1 centers in the diamond sample for the case of the NV center.

\begin{figure*}[t] 
    \centering
    \includegraphics[width=1.0\textwidth]{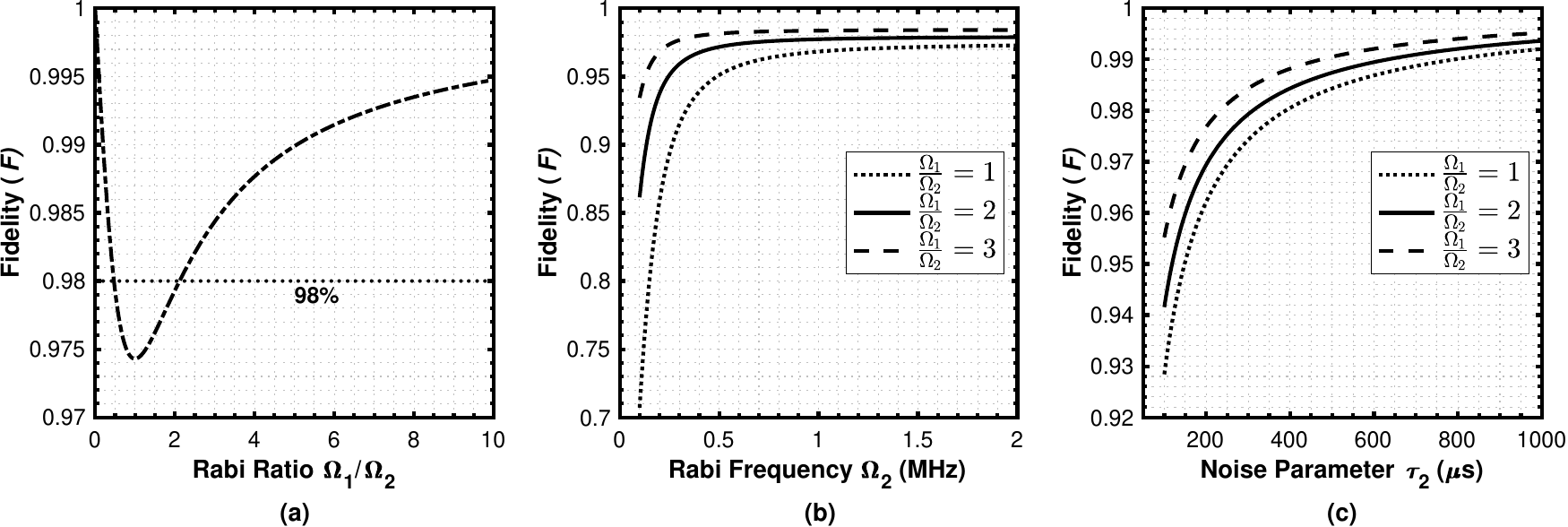}
    \vspace*{-3mm}
    \caption{(Color online) 
    (a) The fidelity of the initialized state as a function of the Rabi ratio for a fixed $\Omega_{2}=10$MHz and noise parameter $\tau_{2}=300\mu s$. The dip is associated with maximum coupling between the dark and bright states. (b) The fidelity of the initialized state as a function of $\Omega_{2}$ for a fixed Rabi ratio and noise parameter $\tau_{2}=300\mu s$. (c) The fidelity of the initialized state as a function of noise parameter $\tau_{2}$ for a fixed $\Omega_{2}=10$MHz and Rabi ratio. Unit fidelity implies that the initialization is at the target dark state. The vacuum decay rate in all three plots is fixed at $\Gamma=1$MHz. }
    \label{figure3}
\end{figure*}

In Fig.~(\ref{figure2}b), we present the dependence of the height of the CPT dip on the noise parameter for the special case of an exponentially decaying correlation function as considered before.
Fig.~(\ref{figure2}b) allows us to infer the value of $\tau_{2}$ (equivalently, $\alpha$) given the CPT simulation data in Fig.~(\ref{figure2}a) and any {\em a priori} model for $C(\tau)$. 
This ability to determine $\alpha$ from an experimentally accessible quantity (the noisy CPT photon count) without resorting to multiple experimental set-ups may be especially advantageous in practice. 

It is important to point out that our results are limited by the weak coupling approximation made in Sec.~\ref{sub:tcl}, namely that for a fixed correlation time $\tau_{c}$, we need to have a sufficiently small bath coupling $g << \frac{1}{\tau_{c}}$. Consequently, not all values of the noise parameter $\alpha$ can be estimated using our developed techniques. To find the applicability range, recall that $\alpha=c_{0}^{2}\tau_{c}$, where we adopted $\tau_{c} \sim 1 \mu s$ consistent with the coupling strength $g << 1$MHz satisfied for P1 centers in diamond with concentrations less than 100ppm \cite{simanovskaia2013sidebands}. In conjunction with $\gamma_{e}c_{0}\approx 0.5$MHz \cite{wang2013spin}, this leads to the following bound:  $\gamma_{e}^{2}\alpha=\gamma_{e}^{2}c_{0}^{2}\tau_{c}<<\gamma_{e}^{2}c_{0}^{2}/g \approx 2.5$MHz (i.e. $\tau_{2}>>0.4\mu$s in Fig.~(\ref{figure2}b)) for $g\approx 0.1$MHz.


\subsection{Qubit state preparation}
\label{applications2}

State initialization of a $\Lambda$ system in an arbitrary superposition of its ground states has been studied \cite{yale2013all}, with application in optically controlled solid-state spin-based quantum devices. This is accomplished by properly tuning the Rabi frequencies $\Omega_{1}, \Omega_{2}$, as well as the phase difference $\phi_{1}-\phi_{2}$ between the two driving fields, 
and using CPT to initialize the system in the dark state
\begin{equation}
|d\rangle =\cos{\theta}|1\rangle-e^{i\phi}\sin{\theta}|2\rangle ,
\label{eqn:dark_state}
\end{equation}
where $\tan{\theta}=\Omega_{1}/\Omega_{2}$ and $\phi=\phi_{1}-\phi_{2}$. The dark state is reached when the laser frequencies $\omega_{L1}$, $\omega_{L2}$ are completely in tune with the transition frequencies of the $\Lambda$ system.
Due to the spin-bath noise, the steady-state solution of the ME Eq.~(\ref{eqn:rate}) will not exactly be a dark state; hence we look for its fidelity with respect to $|d\rangle$. This fidelity is a function of the experimental parameters $\Omega_{1}, \Omega_{2}$ and the noise parameter $\tau_{2}$
\begin{equation}
f(\Omega_{1},\Omega_{2}, \tau_{2})\equiv F(\rho^{\text{eq}}, |d\rangle \langle d|)=\langle d|\rho^{\text{eq}}|d\rangle.
\end{equation}
We assume for simplicity that $\phi=0$ (recall that the noiseless limit corresponds to $\tau_{2} \rightarrow \infty$).
Since the Rabi ratio $\Omega_{1}/\Omega_{2}$ determines the dark state in Eq.~(\ref{eqn:dark_state}) when $\phi=0$, we write the fidelity in a more convenient form
\begin{equation}
f(\Omega_{1},\Omega_{2}, \tau_{2})=g({\Omega_{1}}/{\Omega_{2}},\Omega_{2}, \tau_{2}) . 
\label{eqn: fidelity}
\end{equation}
where $g$ is a function that depends on $\Omega_{1}$ only through the Rabi ratio $\Omega_{1}/\Omega_{2}$.

Fig.~(\ref{figure3}a) shows the range of possible target dark states with different Rabi ratios when the parameters $\Omega_{2}$ and $\tau_{2}$ are fixed. A certain threshold on the fidelity (e.g. $F>0.98$)
should be demanded for a state preparation to be successful. The fidelity plot has a dip at around the Rabi ratio $\Omega_{1}/\Omega_{2}=1$, corresponding to the dark state $|d\rangle = (|1\rangle -|2\rangle)/\sqrt{2}$. This is explicitly shown in Appendix~(\ref{appendixA}). Briefly, the dark state (which is decoupled from the bright state when no spin bath is present) gets coupled to the bright state $|b\rangle$, mediated by the fluctuating magnetic field $b(t)$. The coupling constant is given by $(\sin{2\theta})\gamma_{e}$ which is maximized when $\theta=\pi/4$, corresponding to $\Omega_{1}/\Omega_{2}=1$. Since $|b\rangle$ is coupled to the excited state, this will reduce the fidelity of the target state.

Next, we consider a fixed target dark state (i.e. a fixed Rabi ratio in Eq.~(\ref{eqn: fidelity})). Fig.~(\ref{figure3}b) shows the fidelity as a function of $\Omega_{2}$ for a fixed Rabi ratio and noise parameter. We see that the fidelity quickly saturates with increasing $\Omega_{2}$. This is important because lasers in practice have a finite spectrum width around the desired frequency $\omega$. In the presence of other excited states, this might lead to undesired excitation that takes the electrons out of the $\Lambda$ system (the probability of which is proportional to the matter-field coupling, i.e., the Rabi frequency). Hence, Fig.~(\ref{figure3}b) shows that one should pick the smallest $\Omega_{2}$ for which the fidelity of the target state saturates. For the specific $\Lambda$ system under consideration, the undesired excitation to the next allowed level in the NV center (which is $A_{1}$) can be ignored because the energy difference between the excited states $A_{1}$ and $A_{2}$ is about 3GHz, whereas in practice the driving laser frequencies can get to the desired transition energies within an uncertainty of few tens of MHz. Here, it is also worth mentioning that a similar recommendation can be made against the possibility of optical ionization of the NV center, although in such circumstances, one usually applies periodic "repump" pulses to bring the NV center back to its NV$^{-}$ charge state.

Finally, Fig.~(\ref{figure3}c) showcases the fidelity as a function of the noise parameter $\tau_{2}$ for fixed Rabi frequencies $\Omega_{1},\Omega_{2}$. This shows the noise threshold for which we can expect a preparation of the $\Lambda$ system in the ideal dark (qubit) state, with a certain fidelity. Consider, for example, a fixed Rabi ratio of 1 (i.e. the dark state is in an equal superposition of $|+1\rangle$ and $|-1\rangle $ ground states), then the "acceptable" fidelity of $98 \%$ is achieved for diamond samples with noise parameter value $\tau_{2} > 400\mu$s. For values of $\tau_{2}<400 \mu$s, the CPT method of state initialization \cite{yale2013all} fails to accumulate sufficient fidelity for the equal superposition target state, e.g., $F\approx 0.975<0.98$ for $\tau_{2}=300\mu s$. Note that achieving sufficient preparation fidelity also depends on the choice of the target state, as seen in Fig.~(\ref{figure3}a). Therefore, the decision to use the CPT method for state initialization should be accompanied by the knowledge of the noise parameter $\tau_{2}$ to have a sense of the resulting state fidelity. This is accomplished by Figs.~(\ref{figure2}a, \ref{figure2}b) of our results.

\section{Conclusion}

We analyze the CPT phenomenon in the presence of a classical weakly coupled noise environment, in addition to quantum vacuum noise. We derive a TCL ME for the reduced dynamics of the driven $\Lambda$ system and show that the equilibrium state has a non-zero excited state population. We find a one-to-one correspondence between the height of the CPT dip and the value of the unknown noise parameter, allowing for the determination of the noise without resorting to multiple experimental set-ups. We illustrate our approach by tackling the problem of dissipative qubit state initialization and show that the target states prepared via dark state initialization generally vary in fidelity, with the minimum at the equal superposition state. 

Our work adds upon various related fields, such as quantum sensing, quantum metrology, quantum information processing (e.g. decoherence free subspaces, etc.), and quantum computing. For example, Sec.~\ref{applications2}. in our article sheds light on how well various pure qubit states can be prepared in the laboratory as dark states of a $\Lambda$ system \cite{yale2013all} and what fidelities to expect as a function of the desired amplitudes of $|+1\rangle$ and $|-1\rangle$ qubit states. This is useful if we need to identify the error bars for a quantum information processing, or quantum computation task, based on an imperfect qubit state preparation. Moreover, our application in Sec.~\ref{applications1} could be implemented in quantum sensing of magnetic fields in NV centers (complementing e.g. \cite{wu2021continuous}), as well as quantum metrology, where the goal is to estimate an unknown parameter of a noisy channel. Furthermore, the developed theoretical method for dealing with two baths (classical and quantum) in Sec.~\ref{sub:tcl} can prove useful in various important setups in AMO physics of driven systems, such as when the system is driven by electromagnetic field but is also subject to noise from its physical environment. Finally, in fault-tolerant quantum computing, our parametric noise estimation in Sec.~\ref{applications1} can be done in real time \cite{wu2021continuous}, which means that we can keep track of changes in the environment noise parameter by detecting such changes in the real-time dark counts of our $\Lambda$ system. This makes our three-level system a useful ``spectator'' system in such architectures, by aiding in feedback driven quantum control of computational qubits, see e.g. \cite{majumder2020real}.

Future efforts will be directed towards employing this noise parameter estimation method to monitor an environment with non-stationary noise. Consequently, this provides additional corrective information for the state of a nearby qubit against noise where quantum control can be applied in a feedback loop to maintain high fidelity of the qubit state.

\section*{Acknowledgement}
A.D., A.M., P.B., and N.B. would all like to dedicate this paper to the memory of their advisor, Jonathan P. Dowling; may he rest in peace. We would like to acknowledge Lorenza Viola and Leigh M. Norris for their guidance and support throughout the project as well as their important contribution in the theoretical development of the paper. We would also like to thank Hailin Wang, Shu-Hao Wu, and Ethan Turner for useful discussions. A.D. would like to thank Lorenza Viola for her hospitality during his visit to Dartmouth College. A.D. would also like to thank Vishal Katariya, Mark M. Wilde, Lior Cohen, and Hwang Lee for suggestions and comments. This work was supported by the U.S. Army Research Office through the U.S. MURI Grant No. W911NF-18-1-0218. 

\begin{appendix}
\section{Stochastic Hamiltonian in the dark and bright state basis}
\label{appendixA}

Here we derive the Hamiltonian of a driven $\Lambda$ system in the dark-bright-common (dbc) basis \cite{shakhmuratov2004dark}. First, we write the $\Lambda$ system Hamiltonian
$H(t)=H^0_{\Lambda}+H_{\text{drive}}(t)+H_{c}[b(t)]$ in the presence of the stochastic bath as
\begin{equation}
    \begin{aligned}
        H(t) &=\sum_{n=0,1,2}\omega_{n}P_{nn} -\gamma_{e}b(t)(P_{11}-P_{22})  \\&+\frac{\Omega_{1}}{2}(P_{10}e^{-i(\omega_{L1}t+\phi_{1})} +P_{01}e^{i(\omega_{L1}t +\phi_{1})}) \\&+\frac{\Omega_{2}}{2}(P_{20}e^{-i(\omega_{2L}t+\phi_{2})}+P_{02}e^{i(\omega_{L2}t+\phi_{2})}),
    \end{aligned}
\end{equation}
where $P_{ij}=|i\rangle \langle j|$ are one-dimensional projectors. Next, we transform the wavefunction using the unitary 
\begin{equation}
    U_{\Lambda}(t)=e^{i\sum^{2}_{n=0}\omega_{n}P_{nn}t } \hspace{0.25cm} \Rightarrow \hspace{0.25cm} |\Phi(t)\rangle =U_{\Lambda}(t)|\Psi(t)\rangle ,
\end{equation}
so that the new wavefunction satisfies 
\begin{equation}
    \begin{aligned}
        i\partial_{t}|\Phi(t)\rangle &=i(\partial_{t}U_{\Lambda}(t))|\Psi(t)\rangle+U_{\Lambda}(t)(i\partial_{t}|\Psi(t)\rangle) \\
        &=H_{\text{eff}}(t)|\Phi(t)\rangle ,
    \end{aligned}
\end{equation}
with the effective Hamiltonian given by 
\begin{equation}
\begin{aligned}
    H_{\text{eff}}(t)  
    =&-\gamma_{e}b(t)(\hat{P}_{11}-\hat{P}_{22})\\&+\frac{\Omega_{1}}{2}\hat{P}_{10}e^{-i\phi_{1}} +\frac{\Omega_{2}}{2}\hat{P}_{20} 
    e^{-i\phi_{2}}+\text{H.c.} 
    \label{eqn:effective_hamiltonian}
\end{aligned}
\end{equation}
Notice that we used the rotating basis $|\hat{1}(t)\rangle =e^{-i(\omega_{10}-\omega_{L1})t}|1\rangle $, $|\hat{2}(t)\rangle =e^{-i(\omega_{20}-\omega_{L2})t}|2\rangle $, and $|\hat{0}\rangle=|0\rangle$ to define the new projectors $\hat{P}_{ij}$ (also note that $\hat{P}_{ii}=P_{ii}$).\\
The next step is to move to the dbc-basis, that is, 
\begin{equation}
\begin{aligned}
    |c\rangle &=|\hat{0}\rangle ,\\
    |d\rangle &=e^{i\phi_{2}}\cos{\theta}|\hat{1}\rangle-e^{i\phi_{1}}\sin{\theta}|\hat{2}\rangle ,\\
    |b\rangle  &=e^{-i\phi_{1}}\sin{\theta} |\hat{1}\rangle+e^{-i\phi_{2}}\cos{\theta}|\hat{2}\rangle ,
\end{aligned}
\end{equation}
where, as in the main text, $\tan{\theta}=\Omega_{1}/\Omega_{2}$.\\
We now show that, in the absence of the classical noise $b(t)$, the $\Lambda$ system can be thought of as a single decoupled state $|d\rangle$ and an effective driven two-level system given by the other two states ($|b\rangle$ and $|c\rangle$), with an effective Rabi frequency of $\Omega=\sqrt{\Omega^{2}_{1}+\Omega^{2}_{2}}$. To start, we write 
\begin{equation}
\begin{aligned}
    |\hat{1}\rangle &=e^{-i\phi_{2}}\cos{\theta}|d\rangle+e^{i\phi_{1}}\sin{\theta}|b\rangle ,\\
    |\hat{2}\rangle  &=-e^{-i\phi_{1}}\sin{\theta} |d\rangle+e^{i\phi_{2}}\cos{\theta}|b\rangle , 
\end{aligned}
\end{equation}
which gives 
\begin{equation}
\begin{aligned}
    \hat{P}_{10} &=e^{-i\phi_{2}}\cos{\theta}P_{dc}+e^{i\phi_{1}}\sin{\theta}P_{bc}, \\
    \hat{P}_{20}  &=-e^{-i\phi_{1}}\sin{\theta} P_{dc}+e^{i\phi_{2}}\cos{\theta}P_{bc}.
\end{aligned}
\end{equation}
Substituting into Eq.~(\ref{eqn:effective_hamiltonian}), we find $\Omega(P_{bc}+P_{cb})/2$ for the drive contribution, and the dark state is decoupled from the other two states, as claimed. After including the stochastic contribution of Eq.~(\ref{eqn:effective_hamiltonian}) and using the relationships 
\begin{equation}
    \begin{split}
        P_{11} &=\cos^{2}{\theta}P_{dd}+\sin^{2}{\theta}P_{bb}\\& \qquad +\sin{\theta}\cos{\theta}(e^{-i(\phi_{1}+\phi_{2})}P_{db}+\text{H.c.}), \\
        P_{22} &=\sin^{2}{\theta}P_{dd}+\cos^{2}{\theta}P_{bb}\\
        & \qquad-\sin{\theta}\cos{\theta}(e^{-i(\phi_{1}+\phi_{2})}P_{db}+\text{H.c.}), 
    \end{split}
\end{equation}
we find 
\begin{equation}
\begin{aligned}
    P_{11}-P_{22}
         &= \cos{2\theta}(P_{dd}-P_{bb}) \\ 
         &+ e^{-i(\phi_{1}+\phi_{2})}\sin{2\theta}P_{db} +\text{H.c.}
\end{aligned}
\end{equation}
Therefore, the effective Hamiltonian finally reads
\begin{equation}
\begin{aligned}
    H_{\text{eff}}(t) &= -\gamma_{e}\cos{2\theta}b(t)(P_{dd}-P_{bb})\\
         &\qquad -\gamma_{e}\sin{2\theta}b(t)(e^{-i(\phi_{1}+\phi_{2})}P_{db}
         +\text{H.c.})
         \\&\qquad  \qquad  +\frac{\Omega}{2}(P_{bc}+P_{cb}),
\end{aligned}
\end{equation}
where the first term 
describes the coupling of the dark and bright states to the stochastic magnetic field $b(t)$ with a strength that depends on the ratio of the two Rabi frequencies via $\cos{2\theta}$. The second term describes a coupling between the dark and bright states mediated by the stochastic magnetic field, with a strength that is also determined by the the ratio of the two Rabi frequencies via $\sin{2\theta}$. Finally, the term in the last line 
is the well known coupling between the bright and common states (which does not include the dark state).

When $\sin{2\theta}=0$, we expect the steady-state solution of the ME given in Eq.~(\ref{eqn:rate}) of the main text to have the highest fidelity because the dark and bright states are uncoupled for $\sin{2\theta}=0$. This is the case when $\theta=0 $ (i.e., $\Omega_{1}=0$) or $\theta=\frac{\pi}{2}$ (i.e., $\Omega_{2}=0$). On the other hand, the coupling between the dark and bright states is maximized (and hence the fidelity of the steady state is minimized) when $\sin{2\theta}=1$ (i.e., $\theta=\frac{\pi}{4}$), which corresponds to $\Omega_{1}=\Omega_{2}$. 
This explains the dip in Fig.~(\ref{figure3}a).
\vspace{0.35cm}

\end{appendix}

\bibliographystyle{unsrt}
\bibliography{main.bib}
\end{document}